\documentclass[reqno,12pt]{article}
\usepackage{amsmath}
\usepackage{bm}
\usepackage{a4wide,amssymb}
\usepackage{graphicx,xcolor}
\usepackage{epstopdf}
\usepackage{bbm}
\usepackage[latin9]{inputenc}

 \newcommand{\nn}{\nonumber}
\newcommand{\R}{{\mathbb R}}

\renewcommand{\a}{\alpha}

\renewcommand{\d}{\delta}
           
\newcommand\om{\omega}

\newcommand{\e}{\varepsilon}

\newcommand{\pa}{\partial}

\newcommand{\z}{\zeta}
\newcommand{\la}{\lambda}
\newcommand{\D}{\Delta}
\newcommand{\Lam}{\Lambda}
\def\G{\Gamma}
\def\s{\sigma}

\def\bz{{\bf z}}
\def\bx{{\bf x}}
\def\bv{{\bf v}}

\def\bo{{\bm \o}}

\def\bt{{\bf t}}
\def\bs{{\bm \s}}
\def\bk{{\bf k}}
\def\bze{{\bm \zeta}}
\def\bxi{{\bm \xi}}
\def\bet{{\bm \eta}}

\def\T{{\mathcal T}}
\def\L{{\Lambda}}

\def\RRR{{\mathbb R}}
\def\NNN{{\mathbb N}}

\def\MM{{\cal M}}

\def\CC{{\cal C}}
\def\TT{\mathtt{T}}

\newtheorem{prop}{Proposition}
\newtheorem{thm}{Theorem}

\newtheorem{cor}{Corollary}

\def\be{\begin{equation}}
\def\ee{\end{equation}}
\def\bea{\begin{eqnarray}}
\def\eea{\end{eqnarray}}
\def\ni{\noindent}
\def\nn{\nonumber}

\def\ol{\overline}

\def\d{\delta}
\def\o{\omega}
\def\b{\beta}

\begin{document}


\begin{center} 
{\bf \Large{On the evolution of the empirical measure for the Hard-Sphere dynamics}}

\vspace{1.5cm}
{\large M. Pulvirenti$^{1}$ and S. Simonella$^{2}$}

\vspace{0.5cm}
{$1.$\scshape {\small \ Dipartimento di Matematica, Universit\`a di Roma La Sapienza\\ 
Piazzale Aldo Moro 5, 00185 Roma -- Italy \\ \smallskip
$2.$\ Weierstrass Institute \\ Mohrenstrasse 39, 10117 Berlin -- Germany}}
\end{center}

\vspace{1.5cm}
\noindent
{\bf Abstract.} We prove that the evolution of marginals associated to the empirical measure 
of a finite system of hard spheres is driven by the BBGKY hierarchical expansion. The usual hierarchy 
of equations for $L^1$ measures is obtained as a corollary. We discuss the ambiguities arising in the 
corresponding notion of microscopic series solution to the Boltzmann-Enskog equation.

\vspace{0.5cm}\noindent

%

\vspace{1.5cm}


\section{Introduction} \label{sec:intro}
\setcounter{equation}{0}    
\def\theequation{1.\arabic{equation}}

The hard-sphere dynamics plays a central role in the theory of dynamical systems and as underlying 
microscopic model in kinetic theory. The aim of this paper is to study some features of the time evolution
associated to the finite hard-sphere model, namely we consider $N$ identical hard spheres of diameter $\e$ moving in the whole space $\RRR^3$, H-S system in the sequel. Let 
$$
\bz=\{ z_i\}_{i=1}^N\;, \quad  z_i=(x_i,v_i)\in \RRR^6\;,  \quad  i=1,2, \cdots N\;,
$$ 
be the initial configuration of the system. Here we denote by  $(x_i,v_i)$  the position and velocity of the $i$-th particle.
The evolution is deterministic and given by the usual laws of elastic reflection. Let
$$
\bz (t) =\{ z_i(t)\}_{i=1}^N\;, \quad  z_i (t) =(x_i(t) ,v_i(t) )\in \RRR^6\;,  \quad  i=1,2, \cdots N\;,
$$
be the configuration at time $t$. On a full-measure set of initial configurations, the flow $\bz \rightarrow \bz (t)$
exists for all times \cite{Ale75}.

Choosing an initial configuration $\bz$ for which the dynamics is well posed, one defines the {\em empirical measure} as
\be
\label{emp}
\mu_N (t, dz) = \frac 1N \sum_{i=1}^N \d (z-z_i(t)) \,dz \;,
\ee 
where $\d(\cdot)$ is the Dirac measure at the origin (we drop the dependence of $\mu_N$ on $\bz$).
The sum of Dirac masses is transported through the H-S flow. 
Observe that the evolution of \eqref{emp} is simpler than the 
evolution of, say, $L^1$ measures, since it amounts to follow {\em one} single trajectory. 

Is there a partial differential equation describing the evolution of $\mu_N$?
Of course the empirical measure will not satisfy any closed equation, but rather a {\em hierarchy} of equations
involving the higher order ``empirical marginals'' associated to \eqref{emp},
exactly as in the case of absolutely continuous distributions which evolve through the hard-sphere BBGKY hierarchy.
The latter hierarchy has been studied in detail by several authors \cite{Ce72,IP87,Sp85,Uc88,Si13,GSRT12E}. In contrast with 
the well-known BBGKY holding for smoothly interacting systems, its derivation is delicate because of the
singular character of the H-S flow. Indeed the collision operators appearing in the hierarchy are defined by 
integrals on manifolds of codimension one, while the H-S flow is defined only away from a null-measure set.
This problem is even more delicate when one tries to give a meaning to the hierarchy for singular
measures of type \eqref{emp}, since the integration of delta functions is performed on the boundary of the 
space where the H-S flow takes place.

In this paper, we show how to establish and rigorously 
justify the H-S hierarchy for empirical marginals. To do this, we make use of a {\em series solution} 
representation as introduced first in \cite{La75}, where any integration over boundaries is carefully avoided.
In particular, we show that among all the terms appearing in such representation, the ``real'' H-S trajectory is singled out by a mechanism of cancellations.

As in the case of non-singular measures, the result can be obtained with several different methods.
We point out that the method of reference \cite{Si13} does not require any regularity property and therefore
can be easily extended to a general measure including singular parts. Approximation 
with smooth measures can be also used, together with some detailed information on the H-S flow.
In the present paper we shall use a different approach, which is simpler and more natural in 
the case of empirical marginals. The ingredients of the proof are the semigroup property of the flow and
the boundedness of the number of collisions. Moreover, once established the result for empirical marginals, 
the validity of the H-S BBGKY hierarchy for absolutely continuous measures of class $L^1$
can be recovered by computing 
expectation values. Hence, the method of this paper can be also seen as an alternative approach to the validation 
issue discussed in the quoted references.

Connected with the above analysis is  the problem of describing the so called {\em microscopic solutions}
to the Enskog equation, on which we comment in the last part of the paper. In the literature, several versions
of the Enskog equation may be found. Here we will consider the form sometimes called Boltzmann-Enskog 
equation. This is a kinetic equation in which, in contrast with the Boltzmann equation, the diameter $\e$ of the 
particles enters in the expression of the collision operator. More precisely, it reads as
\bea
&& (\pa_t+v\cdot \nabla_x)f(x,v,t) = \la^{-1} \int_{\RRR^3\times S^2_+} dv_1 d\o \ (v-v_1)\cdot\o  \label{EE} \\
&&\ \ \ \ \ \ \ \ \ \ \ \ \ \ \ \ 
\times \Big\{f(x-\om\e,v_1',t)f(x,v',t)- f(x+\om\e,v_1,t)f(x,v,t)\Big\}\;, \nn
\eea
where the unknown $f$ denotes the probability distribution of a given particle. As usual $x$, $v$ and $t$ denote position, velocity and time respectively.
Moreover $S_+^2=\{\om \in S^2 |\ (v-v_1)\cdot\o \geq 0\},$ $S^2$ is the unit sphere in $\RRR^3$ 
(with surface measure $d\o$), $(v,v_1)$ is a pair of 
velocities in incoming collision configuration   and $(v',v_1')$ is the corresponding pair of outgoing 
velocities defined by the elastic reflection rules
\be
\begin{cases}
\displaystyle v'=v-\om [\om\cdot(v-v_1)] \\
\displaystyle  v_1'=v_1+\om[\om\cdot(v-v_1)]
\end{cases} \;.
\label{eq:coll}
\ee
Finally $\la$ is the mean free path.
Note that the Boltzmann equation is recovered, at least formally, for $\e=0$.
In this paper, we will keep fixed $\e>0$.

In 1975 N.N. Bogolyubov \cite{Bog75} observed that there exist solutions to eq. \eqref {EE}
of the form \eqref{emp}, where $\{ z_i (t) \}_{i=1}^N$ denotes the evolution of an $N-$particle H-S system.   
These are called microscopic solutions. 

Let us add here a comment concerning this terminology. 
We recall that, for $N$ large and corresponding small $\e$,
the \eqref{emp} should provide the normalized density of particles in the microscopic description 
of the gas evolving along \eqref{EE}.
Considering $\bz$ as a random variable and under a suitable ``chaotic'' assumption, 
\eqref{emp} can be actually shown to be very close to the solution of the Boltzmann-Enskog equation \cite{PS14}.
However the convergence takes place only in the continuum limit where $N \to \infty$ as $\e^{-2}$.

Bogolyubov's statement concerns instead \eqref{emp} and \eqref{EE} for finite $N$ and $\e$. The
result may look surprising, the Enskog equation being a genuine irreversible kinetic 
equation. To see this, one may introduce the free-energy functional 
$H(f)=\int f \log f \,dx \,dv + \frac 12 \lambda^{-1}\int_{\R^3}  dx \int_{ B(x,\e)}  dy \,  \rho(x)\, \rho(y) $\;,
where $\rho$ is the spatial density and $B(x,\e)$ is the  ball around $x$ and of radius $\e$. 
Then it turns out that $H$ decreases if $f$ is a solution to \eqref{EE},
see e.g. \cite{AC90}. But since the functional $H$ does not make sense on solutions of the type \eqref{emp},
there is no {\em a priori} contradiction. 

As in the case of the H-S hierarchy, the Boltzmann-Enskog collision operator appearing
in the right hand side of \eqref {EE} is not well defined when evaluated in $f=\mu_N$, so that
a discussion on the precise mathematical meaning of the microscopic solutions of \cite{Bog75}
is required. In \cite{Tru12,Tru14} a suitable regularization of $\mu_N$ 
has been used to give a sense to \eqref{EE} in terms of a limiting procedure.
In this paper, we will approach the problem in a different way. 

Motivated by the fact that the concept of series solution
appears to be convenient to justify the microscopic solutions to the H-S hierarchy, we shall
focus on the same notion of solution for the Boltzmann-Enskog equation and on the comparison 
between them. We find it interesting to observe that such a comparison is rather non-trivial. 
The series solution of the H-S system provides a unique well defined result, when applied 
to microscopic initial data. In contrast, the corresponding expansion for the Boltzmann-Enskog 
equation does not allow to avoid the integration of singular measures over boundaries. As a consequence,
not only the single term of the expansion depends on the regularization, but the series is not even
absolutely convergent for short times. The origin of the ambiguity is that the space of singular solutions 
of the kinetic hierarchy allows {\em contractions}, i.e. different particles having the same configuration. 
Unfortunately, a natural prescription preventing this phenomenon seems to be missing.

\smallskip

The plan of the paper is the following. In the next section we introduce  the H-S and Enskog hierarchies 
 for $L^1$ data, together with the tree expansion, which is a basic tool for our analysis.
In Section \ref{sec:MS} we introduce the microscopic states, namely states which are concentrated on a 
single configuration, and extend the notion of hierarchy to this context. After that, 
we analyze the series solution to the Boltzmann-Enskog equation for microscopic initial data.

\section{Hierarchies} \label{sec:hs}
\setcounter{equation}{0}    
\def\theequation{2.\arabic{equation}}

In Section \ref{sec:hs1} below we introduce the H-S system, recall the preliminary results on the H-S dynamics
and explain how to describe the evolution of a class of absolutely continuous measures,
by means of a hierarchy of equations similar to the usual BBKGY hierarchy for smooth potentials. 
An analogous description is also given for the Boltzmann-Enskog evolution. In both cases,
we provide in Section \ref{sec:hs2} the explicit representation of the series solution in terms of the tree 
expansion and of a class of special flows of particles evolving backwards in time. We mainly follow Sec. 2 of ref. \cite{PS14}
in this part. Finally in Section \ref{sec:hs3} we show how to properly formulate the series solution in order
to extend it to the case of singular measures. This requires a discussion on the invertibility properties of the flows.

\subsection{Preliminary results} \label{sec:hs1}
\setcounter{equation}{0}    
\def\theequation{2.1.\arabic{equation}}

We consider a {\em system of $N$ hard spheres} of unit mass and of diameter $\e >0$ moving in $\RRR^3$. 
With
$
z_i=(x_i,v_i)\in \RRR^6
$ 
we indicate the state of the $i$--th particle, $i=1,2, \cdots$, while 
for groups of particles we use the notation $\bz_j = (z_1,\cdots,z_j)\;.$
``Particle $i$'' is a particle whose configuration is labelled by the index $i$.

The particle configuration lives in the {\em phase space} of the system, defined as the subset 
of $\RRR^{6N}$ in which any pair of particles cannot overlap, namely
\be
\MM_N = \Big\{ \bz_N\in\RRR^{6N} \ \Big|\ |x_i- x_k| > 0,\  i,k =1 \cdots j,\  k\neq i \Big\}\;.
\ee

Given a time--zero configuration $\bz_N \in \MM_N$,
we introduce the flow of the $N$--particle dynamics (equations of motion) 
\be
t \mapsto \TT_N (t)\,\bz_N\;, \label{eq:dyn}
\ee
by means of the following prescription.
Between collisions each particle moves on a straight line with constant velocity.
When two hard spheres collide with positions $x_i, x_j$ at distance $\e$, normalized relative distance 
$$\o = (x_j-x_i)/|x_i-x_j|=(x_i-x_j)/\e \in S^2$$ and incoming velocities $v_i, v_j$ (i.e. $(v_i-v_j)\cdot\o <0$), 
these are instantaneously transformed to outgoing velocities $v'_i, v'_j$ (i.e. $(v_i-v_j)\cdot\o >0$) through 
the relations 
\bea
&&v'_i = v_i - \o[\o\cdot(v_i-v_j)]\;,\nn\\
&&v'_j = v_j + \o[\o\cdot(v_i-v_j)]\;.\label{eq:collpp}
\eea
Such a collision transformation is invertible and preserves the Lebesgue measure on $\RRR^{6}$.
Notice also that the flow $\TT_N(t)$ is piecewise continuous in $t$

The above prescription does not cover all possible situations, e.g. triple collisions (three
or more particles simultaneously at distance $\e$) and grazing collisions ($(v_i-v_j)\cdot\o =0$) are excluded. 
However what is important is that the flow is globally defined, almost everywhere in $\MM_N$.
In fact we have from \cite{Ale75} (see also \cite{MPPP76,CIP94}) that there exists in $\MM_N$ 
a subset, whose complement is a Lebesgue null set, such that, for any $\bz_N$ in the subset, the
mapping \eqref{eq:dyn} is a solution of the equations of motion having $\TT_N(0)\,\bz_N = \bz_N.$ 
Moreover, the shifts along trajectories $t \mapsto \TT_N(t)\,\bz_N$ define a one-parameter group of 
Borel maps on $\MM_N$ which leave the Lebesgue measure invariant. 

In particular, it follows that the number of collisions in any finite interval is finite for almost all initial configurations.
Actually, in our situation one has a stronger result \cite{Ill89,Vas79}:
\begin{prop} \label{cor:oN}
There exists $\ol N \in \NNN$ such that the total number of collisions
in the H-S flow is at most $\ol N$.
\end{prop}
We shall make use of this property in the present paper. It is worth stressing that everything
that follows would automatically apply also to $N$ spheres enclosed
in a region ${\Lambda}\subset \RRR^3$ with elastically reflecting boundaries, as in \cite{Ale75},
in which case the above proposition is generally true in any finite time interval only after
removing a suitable Lebesgue null set (one single particle in a box may undergo infinitely
many collisions in a finite time).

A statistical description of the system is provided by assigning on $\MM_N$ an absolutely continuous probability 
measure with density $W$, initially - and hence at any positive time - symmetric in the exchange of the particles. 
Its time evolution is described by the Liouville equation, which in integral form reads
\be
W(\bz_N,t)=W_0( \TT_N(-t)\, \bz_N)\;,  \label{eq:Liou}
\ee
a.e. in $\MM_N$, where $W_0$ is the assigned initial datum.
The $j$-particle marginals for $j=1,\cdots, N$ are given by
\be
f_j(\bz_j, t) = \int_{S(\bx_j)^{N-j}}dz_{j+1} \cdots dz_{N}\,  W(\bz_{j},z_{j+1}, \cdots , z_{N}, t)\;,\label{eq:marg}
\ee
where $\bz_j \in \MM_j$ and 
\be
S(\bx_j)=\Big\{z= (x,v)\in\RRR^6\ \Big|\ |x-x_k| > \e\ \text{ for all } k=1,\cdots, j\Big\}\;.
\ee
Moreover, to simplify the following notations we shall extend the definition over $\RRR^{6j}$
by
\be
f_j(\bz_j) = 0\;,\ \ \ \ \ \ \ \ \ \ \bz_j\in\RRR^{6j}\setminus \MM_j\;.
\label{eq:extmj}
\ee

The evolution equations for the considered quantities were first  derived formally by Cercignani in 
\cite{Ce72} and take the form ({\em H-S BBGKY hierarchy})
\be
\label{BHHS}
\left(\pa_t + {\cal L}_j \right) f_j (\bz_j,t)
= \e^2(N-j)  \sum_{i=1}^j \int_{S^2\times\RRR^3}d\o \ dv_{j+1}\ B(\o; v_{j+i}-v_i)\ 
f_{j+1} (\bz_j ,x_i+\e\o,v_{j+1},t)\;,
\ee
where ${\cal L}_j$ denotes the generator of the $j-$particle dynamics defined by
\be
{\cal S}^\e_j(t) f_j(\bz_j,\cdot) = 
\begin{cases}
e^{-{\cal  L}_j t} f_j(\bz_j,\cdot)= f_j( \TT_j (-t) \bz_j,\cdot) \ \ \ \ \ \ \ \ \ \ \bz_j\in\MM_j\\
0\ \ \ \ \ \ \ \ \ \ \ \ \ \ \ \ \ \ \ \ \ \ \ \ \ \ \ \ \ \ \ \ \ \ \ \ \ \ \ \ \ \ \ \ \ \ \ \bz_j\in\RRR^{6j}\setminus \MM_j
\end{cases}
\label{eq:jpd}
\ee
and the collision kernel is
\bea
B(\o; v_{j+i}-v_i)=\o \cdot (v_{j+i}-v_i)\;.
\label{eq:Bck}
\eea
Notice that, by virtue of \eqref{eq:extmj}, the integration $d\o$ is restricted over the subset of the sphere
\be
\{ \o\in S^2\, |\, \min_{\ell=1,\cdots,j;\ell\neq i}|x_i+\o\e-x_\ell|>\e\}\;.
\ee

The {\em series solution} of the hierarchy, obtained perturbing the $j-$particle evolution, is 
\bea
 f_j (t) = && \sum_{n = 0}^{N-j} \alpha(N-j,n)  \int_0^t dt_1 \int_0^{t_1}dt_2\cdots\int_0^{t_{n-1}}dt_n \nn\\
&& {\cal S}^\e_j(t-t_1)\CC _{j+1}{\cal S}^\e_{j+1}(t_1-t_2)\cdots 
\CC _{j+n}{\cal S}^\e_{j+n}(t_n) f_{0,j+n}\;, 
\label{eq:BBGexp}
\eea
where $f_{j}(\cdot,0) = f_{0, j}$ is the initial datum (marginals of $W_0$),
\begin{equation}
\alpha (r,n)=r(r-1)\dots (r-n+1)\,\e^{2n}
\end{equation}
($\a(r,0) = 1$), and $\CC_{j+1}$ is the collision operator, given by the sum in the right hand side of Eq. \eqref{BHHS}, i.e.
\bea
&& \CC_{j+1} =\sum_{i=1}^j \CC_{i,j+1} \label{eq:Ecollop}\\
&& \CC_{i,j+1} f_{j+1} (\bz_j,t)= \int_{S^2\times\RRR^3}d\o \ dv_{j+1}\ B(\o; v_{j+i}-v_i)\ 
f_{j+1} (\bz_j ,x_i+\e\o,v_{j+1},t)\;.\nn
\eea
The rigorous validation of formula \eqref{eq:BBGexp} has been discussed in \cite{Sp85,IP87,Si13,GSRT12E}
and can be proved under rather weak assumptions on the absolutely continuous initial measure. An alternative
proof will be presented in Section \ref{sec:MS} (Corollary \ref{cor:MS}).
\smallskip

Finally, we introduce the so called {\em Enskog hierarchy}. 

Let $g$ be a solution to the Boltzmann-Enskog Equation \eqref{EE}.
Then the products
\be
g_j(\bz_j,t):=g(t)^{\otimes j}(\bz_j) 
\label{eq:tfjtdef}
\ee
satisfy 
\be
\left(\pa_t+\sum_{i=1}^j v_i \cdot \nabla_{x_i}\right)g_j = \lambda^{-1}\,\CC_{j+1}g_{j+1}\;.
\ee
This can be easily obtained performing a change of variables $\o \to - \o$ inside the positive part
of the collision operator in \eqref{EE}. The corresponding series solution is:
\bea
&& g_j(t)= \sum_{n\geq 0} \lambda^{-n}\int_0^t dt_1 \int_0^{t_1} dt_2 \cdots \int_0^{t_{n-1}}dt_n \nn\\
&& \ \ \ \ \ \ \ \ \ \ \ \ \ \ \ \ \cdot{\cal S}_j(t-t_1)\CC_{j+1}{\cal S}_{j+1}(t_1-t_2)\cdots\CC_{j+n}{\cal S}_{j+n}(t_n) 
g_{0,j+n}\;,
\label{eq:fjexpE}
\eea
where now ${\cal S}_j(t)$ is the free flow operator, defined as
\be
{\cal S}_j(t)f_j(\bz_j,\cdot) = f_j(x_1-v_1t,v_1,\cdots,x_j-v_jt,v_j,\cdot)\;, \label{eq:ffodef}
\ee
and
\be
g_{0,j} =g_0^{\otimes j} \label{eq:indE}
\ee
is the family of initial data.
Existence and uniqueness for the solutions to the Enskog hierarchy have been discussed in 
\cite{Ark90,AC90,Pul96}.

In spite of their formal similarity, the two expansions   \eqref {eq:BBGexp}
and \eqref{eq:fjexpE} are deeply different as explained in \cite{PS14} and 
as confirmed also by the discussion in Section \ref{sec:MSE} below.

\smallskip

Let us conclude here by establishing a fundamental property of both the series expansions 
introduced above,  namely the {\em semigroup propert}, which is a consequence of the same property
holding for the operator \eqref{eq:jpd} and can be formulated in the following way.  

First we introduce the operator $\T_n (\bz_j,t)$ acting on the marginal of order $j+n$ and
describing the $n-$th term of the expansion 
\eqref {eq:BBGexp}:
\be
 \T_n (\bz_j,t) f_{j+n} = \alpha(N-j,n)  \int_0^t dt_1 \int_0^{t_1}dt_2\cdots\int_0^{t_{n-1}}dt_n
{\cal S}^\e_j(t-t_1)\CC _{j+1}\cdots {\cal S}^\e_{j+n}(t_n) f_{j+n}\;. \label{eq:npTo}
\ee
For reasons that will be clear in the next section, we may call $\T_n (\bz_j,t)$ the {\em $n-$particle
tree operator} for the interacting flow.
Eq. \eqref {eq:BBGexp} can be written
\be
\label{fjNgrad}
f_j (\bz_j, t)= \sum_{n= 0}^{N-j} \T_n (\bz_j,t) f_{j+n}(0)\;.
\ee
If $t=t_1 +t_2$, by simple algebraic manipulations it follows that
\be
\label{semig}
f_j (\bz_j, t)= \sum_{n= 0}^{N-j} \T_n (\bz_j,t_1) f_{j+n}(t_2).
\ee

In a similar way for the Enskog flow we define
the $n-$particle tree operator
\be
\T_n^{E} (\bz_j,t) g_{j+n} = \lambda^{-n} \int_0^t dt_1 \int_0^{t_1}dt_2\cdots\int_0^{t_{n-1}}dt_n
{\cal S}_j(t-t_1)\CC _{j+1}\cdots {\cal S}_{j+n}(t_n) g_{j+n}\;,
\ee
so that
\be
\label{fjN-grad}
g_j (\bz_j, t)= \sum_{n= 0}^{\infty} \T_n^E (\bz_j,t) g_{j+n}(0)
\ee
and there holds
\be
\label{semigE}
g_j (\bz_j, t)= \sum_{n= 0}^{\infty} \T^E_n (\bz_j,t_1) g_{j+n}(t_2)\;.
\ee

\subsection{Tree expansion} \label{sec:hs2}
\setcounter{equation}{0}    
\def\theequation{2.2.\arabic{equation}}

In this section we shall write formulas \eqref{eq:BBGexp} and \eqref{eq:fjexpE} 
in a convenient and more explicit way. We follow \cite {PS14,PSS13}.

\smallskip

Extracting the sums from \eqref{eq:Ecollop} in formula \eqref{eq:BBGexp}, one has
\bea
&& f_j(t) = \sum_{n =0}^{N-j} {\sum_{\bk_n}}^* \alpha (N-j,n)
\int_0^t dt_1 \int_0^{t_1}dt_2\cdots\int_0^{t_{n-1}}dt_n \nn\\
&&\ \ \ \ \ \ \ \ \ \cdot{\cal S}^\e_j (t-t_1)\CC_{k_1,j+1}{\cal S}^\e_{j+1} (t_1-t_2)\cdots 
\CC _{k_n,j+n}{\cal S}^\e_{j+n}(t_n) f_{0,j+n}\;, \label{eq:tfnjexp}
\eea
where
\bea
&& {\sum_{\bk_n}}^* = \sum_{k_1=1}^j \sum_{k_2=1}^{j+1}\cdots \sum_{k_n=1}^{j+n-1}\;. \label{eq:specialsumk}
\eea

To describe the summation rule we introduce a useful notation. The
 {\em $n-$collision, $j-$particle tree}, denoted by $\G(j,n)$, is defined as the collection of integers
$k_1,\cdots,k_n$ that are present in the sum \eqref{eq:specialsumk}, i.e.
\be
k_1\in I_j, k_2 \in I_{j+1}, \cdots, k_n\in I_{j+n-1}\;,\ \ \ \ \ \ \mbox{with\ \ \ \ \ \ $I_s=\{1,2,\cdots,s\}$\;.}
\ee
In this way
\be
{\sum_{\bk_n}}^* = \sum_{\G(j,n)}\;. \label{eq:defsumtrees}
\ee

\begin{figure}[htbp] 
\centering
\includegraphics[width=5in]{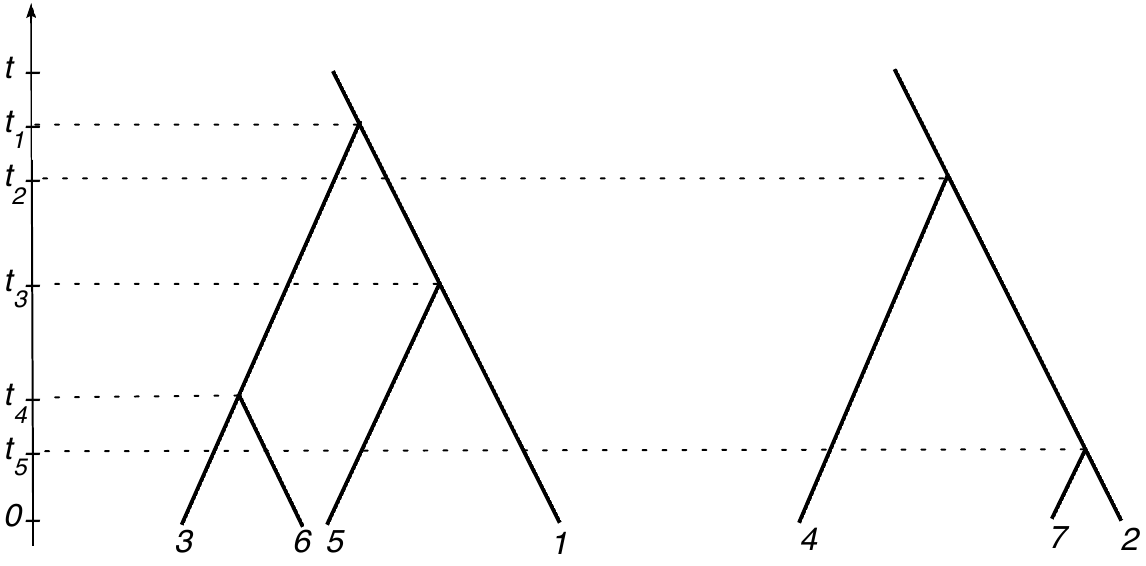} 
\caption{The two--particle tree $\G(2,5)=1, 2, 1, 3, 2.$}
\label{fig:treedef}
\end{figure}

The name ``tree'' is justified by its natural graphical representation,
which we explain by means of an example: see Figure \ref{fig:treedef} corresponding to $\G(2,5)$ 
given by $1, 2, 1, 3, 2.$
In the figure, we have also drawn a time arrow in order to associate times to the nodes of the trees:
at time $t_i$ the line $j+i$ is ``created''. Lines $1$ and $2$ of the example, existing for all times, are called
``root lines''. 

Note that  a $j-$particle tree is not a collection of $j$ one-particle trees because, in the latter case, 
the ordering of particles belonging to different one-particle trees is not specified.

\smallskip
Given a $j$--particle tree  $\G(j,n)$ and fixed a value of all the integration variables in the expansion 
\eqref{eq:tfnjexp} (times, unit vectors, velocities), we associate to it a special 
trajectory of particles, which we call {\em interacting backward flow} (IBF in the following), since it will be 
naturally defined by going backwards in time. The rules for the construction of this evolution are explained as
follows.

\smallskip
First, we introduce a notation for the configuration of particles in the IBF, by making use of Greek alphabet i.e.
$\bze (s)$, where $s \in [0,t]$ is the time\footnote{In previous work \cite{PSS13,PS14} 
the notation $\bze^\e (s)$ has been used for the IBF (and $\bze(s)$ for the corresponding ``Boltzmann flow'').
Here we drop the superscript, since we will keep $\e$ fixed throughout the paper.}. 
Note that there is no label specifying the number of particles. This number
depends indeed on the time. If $s \in (t_{r+1},t_r)$ (with the convention $t_0=t, t_{n+1}=0$), there are exactly $j+r$ 
particles:
\be
\bze (s) = (\z_1(s),\cdots,\z_{j+r}(s)) \in \MM_{j+r}\ \ \ \ \ \mbox{for }s \in (t_{r+1},t_r)\;, 
\label{eq:IBFnot}
\ee
with
\be
\z_i (s)=(\xi_i(s),\eta_i(s))\;,
\ee
the positions and velocities of the particles being respectively
\bea
&& \bxi(s) =(\xi_1(s), \cdots, \xi_{j+r}(s))\;, \nn\\
&& \bet (s)=(\eta_1 (s), \dots, \eta_{j+r} (s))\;. 
\label{eq:IBFnot'}
\eea

Our final goal is to write Eq. \eqref{eq:tfnjexp} in terms of the IBF (to be defined below). More precisely,
\eqref{eq:npTo} shall be rewritten as
\be
\T_n (\bze_j,t)f_{j+n}= \alpha(N-j,n) \sum_{\G(j,n)} \int d {\Lambda}(\bt_n , \bo_n , \bv_{j,n}) \prod_{i=1}^n 
B^\e(\o_i; v_{j+i} - \eta_{k_i} (t_i)) f _{j+n}(\bze (0))\;,
 \label{eq:Teszt}
\ee
where $\bze_j = \bze_j(t) := \bz_j\;,$ $\left(\bze(s)\right)_{s \in [0,t)}$ is defined below,
$d\L$ is the measure on $\RRR^n\times S^{2n}\times\RRR^{3n}$ 
\be
d\Lam ({\bf  t}_n , \bo_n ,\bv_{j,n})= \mathbbm{1}_{\{t_1>t_2 \dots >t_n\}} dt_1\dots dt_n
d\o_1\dots d\o_n dv_{j+1}\dots dv_{j+n}\;,
\ee
with the abbreviation
$$
({\bf  t}_n , \bo_n ,\bv_{j,n})=(t_1, t_2, \cdots, t_n , \o_1, \cdots,  \o_n ,  v_{j+1}, \cdots , v_{j+n} )\;,
$$
while 
\be
B^\e(\o_i; v_{j+i}-\eta_{k_i} (t_i))=B(\o_i; v_{j+i}-\eta_{k_i} (t_i))
\mathbbm{1}_{\{|\xi_{j+i}(t_i)-\xi_{k}(t_i)| > \e\  \forall k\neq k_i\}}
\label{eq:defBe}
\ee
with $B$ defined by \eqref{eq:Bck}.
Summarizing, the term $\T_n (\bze_j,t)f_{j+n}(0)$ is the sum over all trees $\G(j,n)$, of terms where
the initial datum $ f_{0,j+n}$ is integrated, with the suitable weight, over all the possible time-zero states 
of the IBF associated to $\G(j,n)$.

In formula \eqref{eq:Teszt}, the triple $(t_i,\o_i,v_{j+i})$ is thought  as associated 
to the node of $\G(j,n)$ where line $j+i$ is created (see Figure \ref{fig:treedef}). 
In the rest of the paper, we shall further abbreviate 
\be
\int d {\Lambda}(\bt_n , \bo_n , \bv_{j,n}) \prod B^\e = 
\int d {\Lambda}(\bt_n , \bo_n , \bv_{j,n}) \prod_{i=1}^n B^\e(\o_i; v_{j+i} - \eta_{k_i} (t_i)) \;, \label{eq:Tesztbis}
\ee
where the $ \eta_{k_i} (t_i)$ in the factors $B$ have to be computed through the rules specified below, 
starting from the set of variables $(\bt_n , \bo_n , \bv_{j,n})$, the corresponding $j$--particle tree (whose nodes
are labeled by $(\bt_n , \bo_n , \bv_{j,n})$), together with the associated value of $\bze_j, t$.

\smallskip
Let us finally construct $\bze (s)$ for a fixed collection of variables $\G(j,n), \bze_j,\bt_n,\bo_n,\bv_{j,n}$, with
\be
t\equiv t_0 > t_1 > t_2 > \cdots > t_n > t_{n+1}\equiv 0\;, \label{eq:ordertimes}
\ee
and $\bo_n$ satisfying a further constraint that will be specified soon. The root lines of the $j$--particle tree 
are associated to the first $j$ particles, with configuration $\z_1,\cdots,\z_j.$ 
Each branch $j+\ell$ ($\ell=1,\cdots,n$) represents a new 
particle with the same label, and state $\z_{j+\ell}.$ This new particle appears, going backwards in time, at 
time $t_\ell$ in a collision state with a previous particle (branch) $k_\ell\in\{1, \cdots j+\ell -1\},$ with either
incoming or outgoing velocity.

More precisely, in the time interval $(t_r,t_{r-1})$ particles $1,\cdots,j+r-1$ flow according to the usual dynamics
$\TT _{j+r-1}.$ This defines $\bze _{j+r-1}(s)$ starting from $\bze _{j+r-1}(t_{r-1}).$ At time $t_r$ the 
particle $j+r$ is ``created''   by particle $k_r$ in the position
\be
\xi_{j+r} (t_r)= \xi_{k_r} (t_r)+ \o_r \e
 \label{eq:addedpos}
\ee
and with velocity $v_{j+r}$. This defines $\bze (t_r) = (\z_1 (t_r),\cdots,\z _{j+r}(t_r)).$ After that, the evolution
in $(t_{r+1},t_r)$ is contructed applying to this configuration the dynamics $\TT _{j+r}$ (with negative times).

The characteristic function in \eqref{eq:defBe},
is a constraint on $\o_r$ ensuring that two hard spheres cannot be at distance smaller than $\e$.

\begin{figure}[htbp] 
\centering
\includegraphics[width=4in]{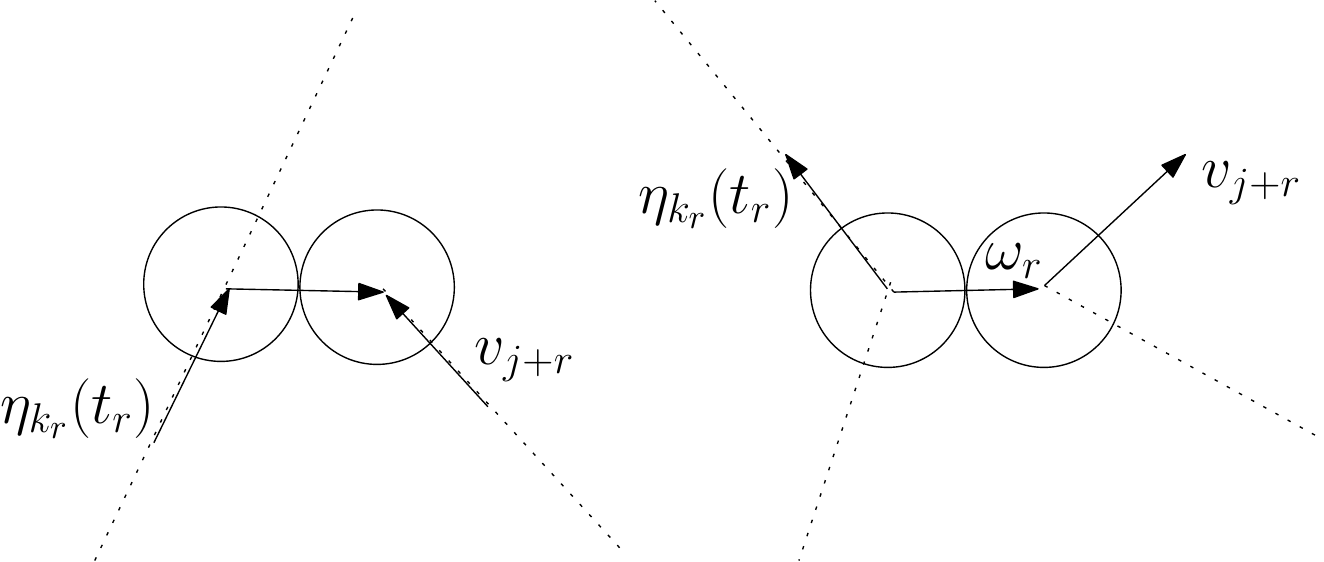} 
\caption{At time $t_r,$ particle $j+r$ is created by particle $k_r,$ either in incoming 
($\o_r \cdot (v_{j+r}-\eta_{k_r} (t_r)) \leq 0$) or in outgoing
($\o_r \cdot (v_{j+r}-\eta_{k_r} (t_r)) \geq 0$) collision configuration. 
Particle $k_r$ is called ``progenitor'' of particle $j+r.$
}
\label{fig:creations}
\end{figure}

We have two cases. If $\o_r \cdot (v_{j+r}-\eta_{k_r} (t_r)) \leq 0$, then the velocities 
are incoming and no scattering occurs, namely after $t_r$ the pair of particles moves backwards freely with velocities 
$\eta _{k_r}(t_r)$ and $v_{j+r}$. If  instead $\o_r \cdot (v_{j+r}-\eta_{k_r}(t_r)) \geq 0$, the 
pair is post--collisional. Then the presence of the interaction in the flow $\TT_{j+r}$ forces the pair to perform a 
(backwards) instantaneous collision. The two situations are depicted in Fig. \ref{fig:creations}.

Proceeding inductively, the IBF is thus constructed for all times $s\in [0,t]$.

\vspace{2mm}
\ni {\bf Remark 1\,\,\,} Between two creation times $t_r, t_{r+1}$ any pair of particles among the existing $j+r,$ different
from the couple $(k_r,j+r),$ can possibly interact. These interactions are called {\em recollisions}, because 
they may involve particles that have already interacted at some creation time (in the future) with another 
particle of the IBF. In our language, recollisions are the ``interactions different from creations''.

\smallskip
To conclude, we have obtained the following representation for the series solution
to the HS-BBGKY hierarchy:
\be
f_j (\bz_j, t)= \sum_{n = 0}^{N-j}  \sum_{\G(j,n)} \alpha (N-j,n)  \int d {\Lambda}(\bt_n , \bo_n , \bv_{j,n}) \prod B^\e\, f _{0,j+n}(\bze (0)) \;.
\label{eq:reds}
\ee

For future purpose, we write also the version of the expansion where we split
positive and negative part of $B$ by setting
$$
B^\e=B^\e_++B^\e_-
$$
with 
$$
B_{\pm}^\e (\o, V) =B^\e  (\o, V) \mathbbm{1}_{ \pm \o\cdot V \geq 0}\;.
$$
If $\bs_n=(\s_1, \cdots, \s_n)$ where $\s_i=\pm$, one has
\be
\label{HS}
f_j (\bz_j , t)= \sum_{n = 0}^{N-j}  \sum_{\G(j,n)} \sum_{\bs_n}  \alpha (N- j,n) \int d {\Lambda}(\bt_n , \bo_n , \bv_{j,n}) \prod B^\e_{\bs_n}\,  
f_{0,j+n}(\bze (0))\;,
\ee
where $\prod B^\e_{\bs_n} = \prod_{i=1}^n B^\e_{\s_i}(\o_i; v_{j+i} - \eta_{k_i} (t_i))$.
We underline that, according to our previous definitions, $B^\e_-$ gives  a negative contribution
to the series expansion \eqref{HS}.

We also notice that the trajectories entering into expansion \eqref{HS} have nothing to do, in principle, with 
the real trajectories performed by the particle system. However the latter can be recovered by the expansion,
by means of a complex system of cancellations. The forthcoming discussion on the hierarchy for empirical measures
will clarify this point.

\smallskip

Finally, a similar analysis can be done for the Enskog hierarchy.
We introduce the backwards flow $\bze^E(s)$, called {\em Enskog backwards flow} (EBF), 
which is constructed as $\bze(s)$ with the additional prescription that {\em all the recollisions are 
ignored}. In this flow, particles are still created at distance $\e$ from their progenitor, but they may 
{\em overlap}, i.e. they may reach a distance smaller than $\e$ during the evolution. In particular,
the time-zero state $\bze^E (0)$ lies in $\RRR^{6(j+n) } $.

Alternatively, we may say that the EBF is constructed exactly as the IBF, except for the following differences:

\ni - the interacting dynamics $\TT$ is replaced by the simple free dynamics;

\ni - there is no constraint on $\o_r$.

In terms of this flow, Eq. \eqref{eq:fjexpE} can be written explicitly
\be
g_j (\bz_j,t)= \sum_{n=0}^\infty \sum _{\G (j,n)} \sum_{\bs_n} \lambda^{-n}
\int d {\Lambda}(\bt_n , \bo_n , \bv_{j,n}) \prod  B_{\bs_n} \,   g_{0,j+n}( \bze^E(0)) \;.
\label{eq:gjbzjt}
\ee
where $\prod B_{\bs_n} = \prod_{i=1}^n B_{\s_i}(\o_i; v_{j+i} - \eta^{E}_{k_i} (t_i))$
and $B_{\pm} (\o, V) =B  (\o, V) \mathbbm{1}_{ \pm \o\cdot V \geq 0}\;.$

It is not difficult to check that the EBF allows a complete factorization, whenever the initial datum does \cite{PS14}.
Namely if $g^{\e}_{0,j}$ is taken as in \eqref{eq:indE}, then \eqref{eq:gjbzjt} gives $g_j(\bz_j,t)=g_1(t)^{\otimes j}(\bz_j) $.

\smallskip

We end this section by summarizing the differences between \eqref{HS} and \eqref{eq:gjbzjt}:
\begin{itemize} 
\item $\sum_{n=0}^{N-j}$ vs.  $\sum_{n\geq0} $ \hfill (total number of particles of the system)\,;
\item $\alpha (N-j,n)$  vs. $\lambda^{-n}$ \hfill (multiplicative coefficient in the expansion)\,;
\item $B^\e_{\bs_n}$ vs. $B_{\bs_n}$ \hfill (overlap of created spheres prevented / allowed)\,;
\item $f_{0,j+n}$ vs. $g_{0,j+n}$ \hfill (initial data, respectively non-tensorized / tensorized)\,;
\item IBF vs. EBF \hfill (backwards flow with recollisions / overlaps).
\end{itemize}
In particular, we observe once more that, in contrast with $g_{0,j}$ which can be taken
as a product state,  $f_{0,j}$ cannot factorize because it must prevent the overlap between different
spheres (condition \eqref{eq:extmj}).

\subsection{Weak formulation of hierarchical expansions} \label{sec:hs3}
\setcounter{equation}{0}    
\def\theequation{2.3.\arabic{equation}}

In this section we discuss a weak formulation of the above introduced series 
solution to the HS hierarchy, suitable to deal with singular measures.

Let us consider the right hand side of \eqref {HS} in the case that the initial data $f_{0,j+n}$ are
replaced by suitable measures, not absolutely continuous. To give a precise meaning to the formula, 
we need to integrate with respect to the initial configuration variables $\bze(0)$. This amounts to study
in some more detail the {\em IBF-map} defined, for fixed $t>0$, $\G(j,n)$ and 
$\bs_n$, by
\be
\label {map}
\bze_j, \bt_n, \bo_n, \bv_{j,n} \longrightarrow \bze_{j+n} (0)\;,
\ee
where $\bze_j \equiv \bz_j$. 
It follows from the properties of the H-S flow that the above transformation is a Borel map almost everywhere defined
over the domain specified in the previous section, with image $A(t,\G,\bs_n)\subset \MM_{j+n}$\footnote{I.e.,
$d\bze_{j+n}-$essentially, $$A(t,\G(j,0),\emptyset)= \MM_{j+n}\;,$$ $$A(t,\G(j,1)=k_1,\s_1)= \Big\{\bze_{j+n+1}(0)\in
\MM_{j+n+1}\,\Big|\,\mbox{``$j+n+1$ and $k_1$ collide in $(0,t)$"}\Big\}$$ and so on.};
see e.g. Lemma~1 in \cite{Si13}, or \cite{Uc88} for a different method of proof.
Moreover, a simple computation shows that the Jacobian determinant of the transformation is
in modulus $\e^{2n}\prod_{i=1}^n |B_{\s_i}(\o_i; v_{j+i} - \eta_{k_i} (t_i))|$, so that the map induces 
the equivalence of measures
\be
d\bze_j\,d{\Lambda}\, \e^{2n}\prod |B^\e_{\bs_n}| = d\bze_{j+n}(0)\;.
\label{eq:eqmeas}
\ee

We construct now the inverse of the map. To do so, we introduce the {\em interacting forward flow}
(IFF) $$\bze_{j+n}(0) \longrightarrow \left(\bze^{F}(s)\right)_{s \in [0,t]}$$ defined in the following way. 
As for the IBF, the IFF has a number of particles which depends on time. At $s=0$, one has $j+n$ particles.
Let us take for such particles a configuration $\bze_{j+n} (0)$ in the image $A$ of \eqref{map} (which depends
on $t$, $\G(j,n)$ and $\bs_n$). We let the configuration evolve forward in time via the H-S dynamics $\TT_{j+n}$
up to the first collision between particle $j+n$ and particle $k_n$ 
($j+n$ is the last one created in the IBF and generated by particle $k_n$
according to the tree $\G (j,n)$). Such an instant of time exists because $\bze_{j+n} (0)$ belongs to $A$.
There are two possibilities.
\begin{itemize}
\item This interaction is a creation. Then particle $j+n$ disappears, while particle $k_n$ interacts or not according to
$\s_n=+$ or $\s_n=-$ respectively. Next, particle $k_n$ evolves together with the other $j+n-2$ particles in the
H-S dynamics $\TT_{j+n-1}$, up to the next collision dictated by $\G(j,n)$, namely the collision between 
particles $j+n-1$ and $k_{n-1}$.
\item This interaction is a recollision. Then we let evolve the system further with the dynamics $\TT_{j+n}$
up to the next contact between  $j+n$ and $k_n$. Clearly, meanwhile other recollisions occur.  If there is no
next contact in $[0,t]$ between  $j+n$ and $k_n$\footnote{More generally, if iteration of the procedure becomes 
impossible.}, we simply eliminate this second option.
\end{itemize}
As we shall see in Section \ref{subsec:hs3}, there are cases in which both options are possible.
Therefore, the iteration of the above procedure generates $M$ different flows $\bze^{F,i}(s)$,
$i = 1,\cdots, M$, with $M$
depending on $\bze_{j+n}(0)$. Of course by Proposition \ref{cor:oN}, there exists $\ol N\in \NNN$ such that
$1\leq M \leq 2^{\ol N}$.

We conclude that (even though locally invertible) the map \eqref{map} is not globally invertible and 
integrating Eq. \eqref{HS} against a bounded continuous function $\phi=\phi(\bze_j): \RRR^{6j}\to \RRR$, 
the following {\em weak formulation of the time-integrated H-S BBGKY hierarchy} is obtained:
\be
\int f_j (t) \phi= \sum_{n = 0}^{N-j}  \sum_{\G(j,n)} \sum_{\bs_n}  \prod_{r=1}^n \s_r \, 
\alpha (N- j,n)\, \e ^{-2n}  \int_{A(t,\G,\bs_n)}  d\bze_{j+n}\,
f_{0,j+n}(\bze_{j+n})\,\sum_{i=1}^{M} \,\phi (\bze^{F,i}_j(t))\;, \label{eq:Wfhe}
\ee
where $M = M(\bze_{j+n}): A \to \NNN$. Note that the latter function is an a.e.-defined step function 
over $\MM_{j+n}$, strongly dependent on the details of the H-S flow. Observe also that the right hand side
of \eqref{eq:Wfhe} makes sense even for initial marginals which are measures $df_{0,j}(\bze_{j})$.

Though the above formula is not very handable for practical purposes, we shall use it only in the
case of discrete measures. Remind that by \eqref{eq:eqmeas}, the transformation \eqref{map} is 
nonsingular out of the grazing collisions $B_{\s_i}=0$. Therefore for any Dirac measure supported 
in points of $A(t,\G,\bs_n)$ ``avoiding" grazing collisions, \eqref{eq:Wfhe} can
be rewritten without ambiguity as
\be
\int f_j (t) \phi= \sum_{n = 0}^{N-j}  \sum_{\G(j,n)} \sum_{\bs_n}  \alpha (N- j,n) \int d\bze_j\,d{\Lambda}\, \prod B^\e_{\bs_n} \,
f_{0,j+n}(\bze_{j+n} (0))\,\phi(\bze_j)\;. \label{eq:WSfhe}
\ee
In other words, we describe the discrete measure as a pushforward measure along the mapping \eqref{map}.

In Section \ref{sec:MS} we will show that the empirical marginals associated to \eqref{emp} and supported
on ``good" configurations of the H-S flow satisfy indeed \eqref{eq:Wfhe}. We shall call these solutions the 
{\em microscopic solutions} of the 
H-S hierarchy. An obvious way to proceed would be by taking a sequence of data $(f_{0,j})_k$ approximating
the singular measure as $k \to \infty$ and using some topological information on the function $M = M(\bze_{j+n})$. 
Instead, we shall present in Section \ref{sec:MS} a 
more constructive approach. 
 
 \smallskip
 
We conclude by observing that the {\em EBF-map} defined, for fixed $t>0$, $\G(j,n)$ and $\bs_n$, by
\be
\label {mapE}
\bze^{E}_j, \bt_n, \bo_n, \bv_{j,n} \longrightarrow \bze^{E}_{j+n} (0)
\ee
with $\bze^{E}_j \equiv \bz_j$, is globally invertible over its domain of definition. This follows
immediately from the fact that each pair of particles in the EBF may interact at most once.
As a consequence, from \eqref{eq:gjbzjt} we easily deduce the following {\em weak
formulation of the time-integrated Enskog hierarchy}:
\be
\int g_j (t) \phi= \sum_{n = 0}^{\infty}  \sum_{\G(j,n)} \sum_{\bs_n} \prod_{r=1}^n \s_r \,  \lambda^{-n}\, \e ^{-2n}  
\int_{A^{E}(t,\G,\bs_n)}  d\bze^{E}_{j+n}\,g_{0,j+n}(\bze^{E}_{j+n})\,\phi (\bze^{E}_j(t))\;, \label{eq:WfheE}
\ee
where $A^{E}(t,\G,\bs_n)\subset \RRR^{6(j+n)}$ is the image of \eqref{mapE}.
In Section \ref{sec:MSE} we show that this formula is not well-defined for the empirical measure,
since ambiguities arise from the contractions in $g_{0,j+n}$ and the corresponding integral 
of the singular measure over the boundary of $A^{E}$.

\subsubsection{Non-invertibility of the IBF} \label{subsec:hs3}
\setcounter{equation}{0}    
\def\theequation{2.3.1\arabic{equation}}

We show the non-invertibility of the map \eqref{map} by means of an example.

Let us consider the case of $N = 3$ hard spheres, with initial configuration 
$\bze_{3}(0)$ such that the dynamics admits the sequence of collisions as in 
the figure that follows, i.e. $c_1 = (2,3), c_2 = (1,2), c_3 = (2,3), c_4 = (1,2)$. 
\be
\includegraphics[width=2in]{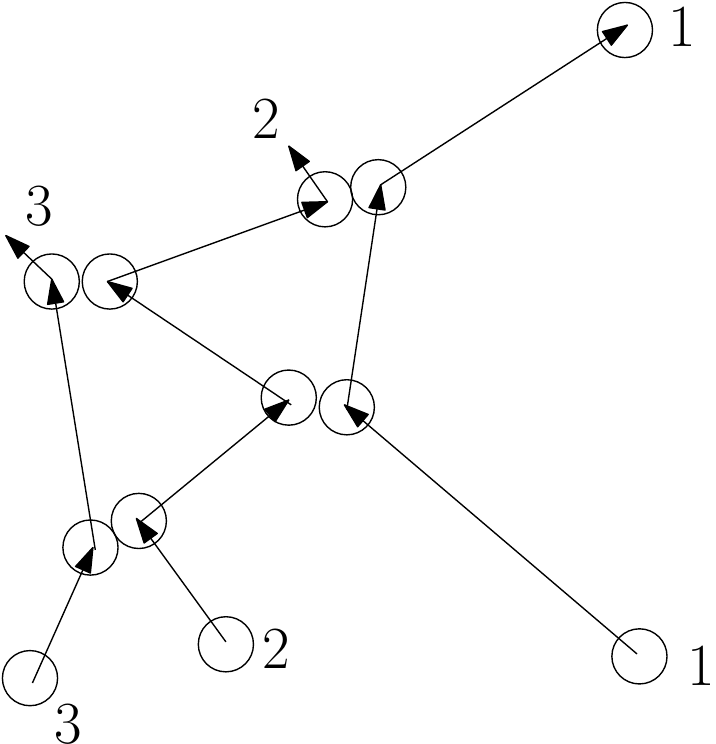}  \label{eq:figI1}
\ee
Many configurations of this type can be constructed in two or three dimensions \cite{MC92}.
The above figure is an IBF for the following tree $\G(1,2)$ with positive nodes $\bs_2=(+,+)$:
\be
\includegraphics[width=2in]{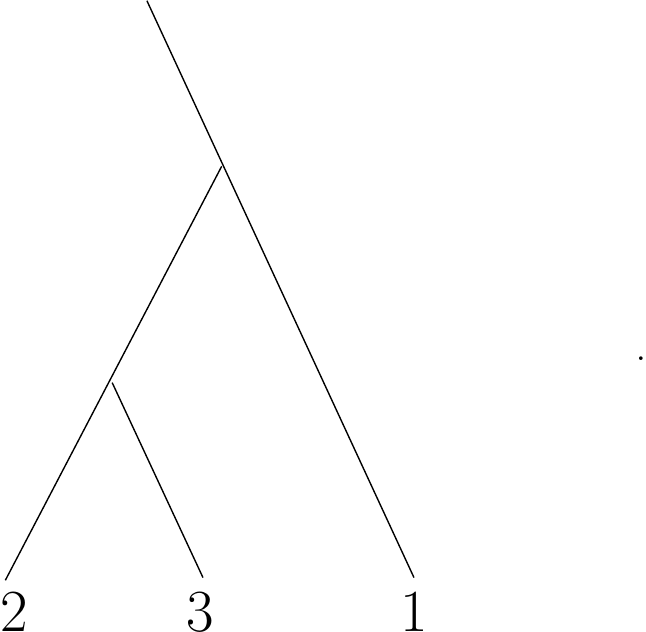} \label{eq:figI2}
\ee
The collisions $c_1$ and $c_2$ correspond to the nodes of the tree, while $c_3$ and $c_4$ are 
recollisions. 

A different IBF for the same $\G(1,2)$ and $\bs_2$ is obtained by taking $c_3$ and $c_4$
as creation times, as in the figure below which, for the same initial configuration and tree, 
yields a different sequence $\z_1,\bt_2, \bo_2, \bv_{1,2}$.
\be
\includegraphics[width=2in]{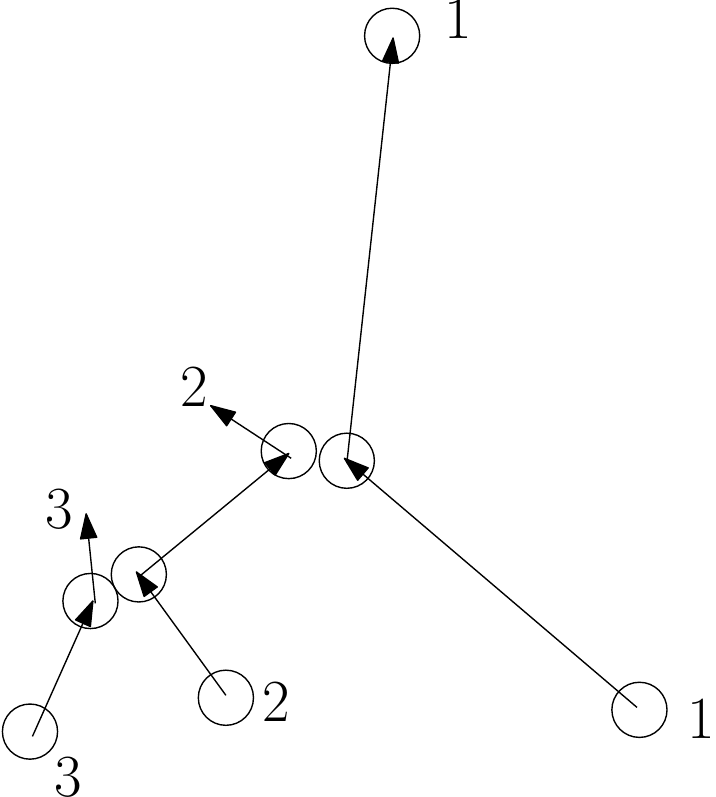} \label{eq:figI3}
\ee


\section{Microscopic solutions to the H-S hierarchy} \label{sec:MS}
\setcounter{equation}{0}    
\def\theequation{3.\arabic{equation}}

Let us start by introducing the probability measure over $\MM_N$ having as $1-$particle marginal
the empirical measure. The density $W$ and marginals $f_j$ introduced in Section \ref{sec:hs1}
for the absolutely continuous case are here replaced by the {\em empirical marginals}, which are 
higher order probability measures concentrated on subsets  of the H-S configuration.

Let $\bz_N = (z_1,\cdots,z_N) \in \MM_N$ be a given configuration. A symmetric probability
measure over $\MM_N$ concentrated on $\bz_N$ is given by the {\em empirical density} 
\begin{equation}
\D (\bze_N)= \frac 1 {N!} \sum_{\pi} \prod _{i=1}^N \d (\z_i- z_{\pi (i)})\;, \label{eq:empd}
\end{equation}
where ${\pi}$  is the generic permutation.

The empirical marginals $\D_j (\bze_j)$ are defined over $\MM_j$, as usual,  by
\begin{equation}
\D_j  (\bze_j)= \int d\z_{j+1} \cdots d\z_{N}\, \D (\bze_j,\z_{j+1}, \cdots, \z_{N})\;.
\end{equation}
It follows that
\begin{align}
\D_1 (\z_1)& = \int d\z_{2} \cdots d\z_{N} \,\D (\z_1,\z_{2}, \cdots, \z_{N})\nn \\ \nn
&= \frac 1{N!} \sum_{\pi} \int d\z_{2} \cdots d\z_{N} \prod_{i=1}^N \d (\z_i- z_{\pi(i)}) \\ \nn
&= \frac 1{N!} \sum_{\pi}  \d (\z_1-z_{\pi(1)}) \\ 
&= \frac 1{N} \sum_{i=1}^N \d (\z_1-z_i)\;, 
\end{align}
namely $$\D_1 (\z)d\z=\mu_N(d\z) $$ is the empirical measure \eqref{emp} at given time.
By a similar computation:
\begin{align}
\label{micromar}
\D_j (\bze_j)&=\frac 1{N!} \sum_{\pi}  \prod_{i=1}^j \d (\z_i-z_{\pi(i)}) \nn\\ \nn
& =\frac {( N-j)!}{N!} \sum_{\substack{i_1,\cdots,i_j \\ i_a \neq i_b}} \prod_{s=1}^j   \d (\z_s-z_{i_s}) \\ 
&= \frac 1{N(N-1) \cdots (N-j+1)}\sum_{\substack{i_1,\cdots,i_j \\ i_a \neq i_b}} \prod_{s=1}^j   \d (\z_s- z_{i_s})\;. 
\end{align}
Observe that over $\MM_j$ one has the identity:
\begin{equation}
\D_j(\bze_j)= \frac {N^j} {N(N-1) \dots (N-j+1)} \mu_N^{\otimes j} (\bze_j) \label{eq:ident}
\end{equation}

Next we consider the time evolution of the marginals. 

First of all, we need to take care on how to fix 
the initial configuration $\bz_N$. By the results of \cite{Ale75,MPPP76,CIP94} quoted in Section \ref{sec:hs1}, 
one can identify the full Lebesgue-measure set of good H-S configurations 
\be
\MM_N^* \subset \MM_N
\ee
with the set for which triple (or multiple) collisions, grazing collisions and simultaneous collisions\footnote{Namely
two pairs of particles colliding at the same time. This does not lead to an ill-defined dynamics, but is conveniently
removed to simplify the following argument.} are forbidden. 
The evolution of such configurations is defined for all times. The following set is of course still full:
\be
\MM_N^\star = \Big\{\bz_N \in \MM_N^*\,\Big|\,\bz_k\in\MM_k^*\,\,\,\forall \bz_k \subset \bz_N\Big\}\;.
\ee
From now on, we shall fix $\bz_N\in \MM_N^\star$.
This is needed to ensure well defined flows in the right hand side of \eqref{eq:Wfhe}-\eqref{eq:WSfhe}.

By definition,
\begin{equation}
\D(\bze_N,t)=\D(\TT_N (-t)  \bze_N)= \frac 1 {N!} \sum_{\pi} \prod _{i=1}^N \d (\z_i- z_{\pi (i)}(t))
\label{eq:empde}
\end{equation}
and (as in \eqref{micromar})
\be
\label{micromart}
\D_j (\bze_j, t)=\frac 1{N(N-1) \dots (N-j+1)} \sum_{\substack{i_1,\cdots,i_j \\ i_a \neq i_b}} \prod_{s=1}^j   \d (\z_s- z_{i_s} (t) ) \;,
\ee
where here and below we indicate $$\bz_N(t) = \TT_N(t)\bz_N\;.$$

\smallskip

We want to show that the sequence formed by the empirical marginals of the time evolved measure, 
$\{ \D_j (\bze_j,t) \}_{j=1}^N$, satisfies the H-S hierarchy, Eq. \eqref{eq:WSfhe}. 

To do this, we use Proposition \ref{cor:oN} and call $S$ the number of collisions delivered by $\bz_N(\cdot)$
in the time interval $[0,t)$. We partition $[0,t)$ by a sequence of $S+1$ intervals
\be
[0,t)=\bigcup_{i=0}^S [t_i,t_{i+1}) \label{eq:PART}
\ee
where $0=t_0 < t_1 < \cdots < t_{S+1}=t$ and the following properties are satisfied:
\begin{description}
\item[Property 1] In each time interval $ (t_i,t_{i+1})$ with $i=0,\cdots, S-1$, the H-S system delivers 
a single collision at time $\tau_{i+1}$. In the last interval $[t_S,t_{S+1})$ the motion is free.
\item[Property 2] If the collision at time $\tau_{i+1}$ occurs between particles $\a$ and $\b$, then 
$t_{i+1}-\tau_{i+1}$ is chosen so small that the following two different H-S flows are free:
$$s \to \TT_{N-1}(s)\Big(z_1(t_i),\cdots,z_{\a-1}(t_i),z_{\a+1}(t_i),\cdots,z_N(t_i)\Big) \qquad s \in [0,t_{i+1}-t_i)$$
and
$$s \to \TT_{N-1}(s)\Big(z_1(t_i),\cdots,z_{\b-1}(t_i),z_{\b+1}(t_i),\cdots,z_N(t_i)\Big) \qquad s \in [0,t_{i+1}-t_i)\;.$$
\end{description}
Note that Property 2 means that if we remove particle $\a$ (or $\b$) just before the collision instant 
$\tau_{i+1}$, then we see a free flow up to $t_{i+1}$. This will allow to restrict the computation of 
\eqref{eq:WSfhe} to $n=1,2$.

The existence of the above partition is guaranteed by the assumption $\bz_N \in \MM_N^*$ and by the continuity
of the free flow.

Now notice that, in view of the {\em semigroup property} Eq. \eqref{semig}, 
it is enough to prove \eqref{eq:WSfhe} for any time interval in the partition. Therefore in what follows
we shall restrict to a time interval $(0,t)$ in which at most one collision takes place and $t$ is so small that 
Property 2 is fulfilled. Without loss of generality we set $\a=1$, $\b = 2$.

For simplicity of notation, we drop the integration $\int d\bze_j \phi(\bze_j)$ in many of the following formulas.
Furthermore, we abbreviate $\D_j(0) = \D_j$.

Let us consider first the case $j=1$.

The term $n=0$ in the expansion \eqref{eq:WSfhe} is
\bea
\label{free}
\T_0 (\z_1,t) \D_{1} = && {\cal S}_1^\e (t) \D_1(\z_1)\nn\\ \nn
=&& \D_1 (\xi_1-\eta_1t, \eta_1) \\ 
=&& \frac 1N \sum _{i=1}^N \d (\xi_1-(x_i+v_i t)) \d (\eta_1-v_i)\;.
\eea
If $S=0$ (no collision is delivered by $\bz_N(\cdot)$), this reproduces trivially $\D_1(t)$. 
Otherwise if $S=1$ (one single collision occurs), all the terms in the above sum 
contribute to $\D_1(t)$ {\em except} those with $i=1,2$, which therefore must be {\em cancelled} 
by some other terms of the expansion with $n>0$.

If $S=0$, all such terms are easily shown to be zero. Let us assume $S=1$
and consider the term with $n=1$ and $\s_1=+$, namely
\bea
\label{int1}
\T_1^+(\z_1,t)\D_2 && \equiv  (N-1) \e^2 \int d{\Lambda}(t_1,\o, v_2) \, B^\e_+\, \D_2 (\bze_2 (0))\\
&& =\frac {(N-1)} { N (N-1)} \e^2   \int d{\Lambda}(t_1,\o, v_2)\, B^\e_+\nn\\
&&\,\,\,\,\,\,\,\cdot\Big[\d( \z_1 (0) -z_1) 
\d( \z_2 (0) -z_2)+\d( \z_1 (0) -z_2) \d( \z_2 (0) -z_1)\Big]\;. \nn
\eea
Observe that only two of all the terms defining $\D_2$ appear in the right hand side of
the equation, because all other pairs in $\bz_N$ do not interact by the above assumed
properties, so that they fall {\em away} from the image of the IBF-map. 
In order to have other contributions it is then necessary that particle $1$ (or particle $2$), 
ignoring particle $2$ (or particle $1$), collides with some other particle. 
But such events are absent because of Property 2, invoked just to avoid these contributions.

To compute the integral in \eqref{int1} we
proceed as in \eqref{eq:Wfhe}, namely we change variables according to $\z_1,t_1,\o_1,v_2
\to \z_1(0),\z_2(0)$ and use \eqref{eq:eqmeas}:
\be
\int d\z_1 \phi(\z_1)\T_1^+(\z_1,t)\D_2 =  \frac {1}{N} \int d\z_1 \, d\z_2 \,
\phi(\z^F_1(t)) \Big[\d( \z_1 -z_1) \d( \z_2 -z_2)+\d( \z_1 -z_2) \d( \z_2 -z_1)\Big]\;,
\ee
where the integral on the right hand side is now extended over $A(t,\G(1,1),+)$, i.e. on the 
set of couples of particles leading to a collision. Since $(z_1,z_2)$ lies inside this 
set, the integral is equal to $\phi(z_1(t))+\phi(z_2(t))$. Therefore we obtain
\be
\T_1^+(\z_1,t)\D_2= \frac{1}{N} \Big\{ \d ( \z_1-z_1(t)) +\d( \z_1 - z_2 (t)) \Big\}\;.
\ee

With the same computation we get for the term with $n=1$ and $\s_1=-$:
\be
\label{T-}
\T_1^-(\z_1,t)\D_2=-\frac 1N \Big\{ \d (\xi_1-(x_1+v_1 t)) \d (\eta_1-v_1)+ \d (\xi_1-(x_2+v_2 t)) \d (\eta_1-v_2)\Big\}\;.
\ee

The term \eqref{T-}, which is negative, {\em cancels} the two terms in \eqref{free} with $i=1,2$.
Hence we conclude that
\be
\label{concl}
\D_1(\z_1,t) =\T_0 (\z_1,t) \D_{1} + \T_1^+(\z_1,t)\D_2 +\T_1^-(\z_1,t)\D_2\;.
\ee

It remains to show that all the other terms in the expansion \eqref{eq:Wfhe} with $n>1$ 
vanish. This follows immediately from Property 1 and Property 2. In fact, in order to find
a subset of $1+n$ particles of $\bz_N$ lying inside the image of the IBF-flow with $n>1$, 
we would need at least one more collision with respect to those allowed by the properties.
 
 For $j>1$, we proceed in the same way. The term $\T_0 (\bze_j,t) \D_{j}$ is given by
 \be
 \label{freeja}
 {\cal S}^\e_j(t) \D_j (\bze_j)=
\frac 1{N(N-1) \dots (N-j+1)}\sum_{\substack{i_1,\cdots,i_j \\ i_a \neq i_b}} \prod_{s=1}^j   \d(\z_s-z_{i_s}(t))\;.
\ee
As before if $S=1$ and particles $1$ and $2$ collide, then the terms which contribute to $\D_j(t)$ are those with either
$$
(1,2) \subset  \{ i_1 \cdots i_j \}
$$
or
$$
(1,2) \cap \{ i_1 \cdots i_j \}=\emptyset\;.
$$
The other terms, namely those for which $1\in \{ i_1 \cdots i_j \}$ and $2\notin \{ i_1 \cdots i_j \}$, or the 
reverse situation, are exactly compensated by $ \T_1^-(\bze_j ,t)\D_{j+1}$, while the contributions missing 
in \eqref{freeja} to reconstruct $\D_j(t)$ are produced by  $\T_1^+(\bze_j ,t)\D_{j+1}$.
Finally the same arguments used for $j=1$ show that all the terms with $n>1$ are vanishing.

\smallskip

In conclusion, we have proved the following result.
\begin{thm} \label{thm:main}
Let $\bz_N \in \MM_N^\star$. The empirical marginals $\D_j (\bze_j,t)$ defined in \eqref{micromar}
and supported in $\bz_N$ verify, for any $t>0$, the BBGKY expansion \eqref{eq:Wfhe}. Namely,
for any bounded continuous $\phi: \RRR^{6j}\to \RRR$,
\be
\int \D_j (t) \phi= \sum_{n = 0}^{N-j}  \sum_{\G(j,n)} \sum_{\bs_n}\, \alpha (N- j,n) \int d\bze_j\,d{\Lambda}\, \prod B^\e_{\bs_n} \,\D_{j+n}(\bze_{j+n} (0),0)\,\phi(\bze_j)\;. \nn
\ee
\end{thm}

By virtue of  \eqref{eq:ident}, this result can be formulated, in terms of the
empirical measure $\mu_N$ over $\MM_j$, as
\be
\left(\mu_N(t)\right)^{\otimes j} = \sum_{n = 0}^{N-j}  \sum_{\G(j,n)} \sum_{\bs_n}\, (N\e^2)^n \int \,d{\Lambda}\, \prod B^\e_{\bs_n} \,\mu_N^{\otimes (j+n)}(\bze_{j+n} (0))\;, \label{eq:mainA}
\ee
where we set $\mu_N(0) = \mu_N$, or, equivalently:
\be
\int_{\MM_j}\left(\mu_N(t)\right)^{\otimes j} \phi= \sum_{n = 0}^{N-j}  \sum_{\G(j,n)} \sum_{\bs_n}  \prod_{r=1}^n \s_r \, 
N^n\,  \int_{A(t,\G,\bs_n)}  d\bze_{j+n}\,
\mu_N^{\otimes (j+n)}(\bze_{j+n})\,\sum_{i=1}^{M} \,\phi (\bze^{F,i}_j(t))\;. \label{eq:mainB}
\ee

\smallskip

 \ni {\bf Remark 2\,\,\,} The semigroup property has been used as essential ingredient 
 to simplify the above proof. Indeed, even without arguing on the time derivative, we can still
 restrict to such a small interval of time, that only a single collision takes place, in both the $N-$particle 
 dynamics and the dynamics of proper subsets of particles. This allows in turn to compute just
the value of $0-$collision and $1-$collision trees, showing compensations among terms
of type $ \T_0$ and terms of type $ \T_1^-$. 
In contrast, a control of the full expansion in the generic time interval would require to detect more complicated 
cancellations. Some of them have been classified in \cite{Si13}. Our proofs, however, do not
characterize the complete set of compensations leading from the {\em virtual} trajectories
of the expansion to the unique, {\em real} motion of particles appearing on the left hand side.
 
We provide one example of such compensations, in the case of $N=3$ hard spheres 
already considered in Section \ref{subsec:hs3}. The collisions therein called $c_1,c_2,c_3,c_4$ 
are all the collisions exhibited by the dynamics in $[0,t]$; see Figure \eqref{eq:figI1}.
 Let $\D_1(s)$ the empirical distribution 
supported on this trajectory. Then the Dirac mass of particle $1$ at time $t$ is produced in 
$z_1(t)$ by the term $\G(1,2)= (1,2)$ with $\bs_2=(+,+)$ in the BBGKY expansion (i.e. the tree 
pictured in \eqref{eq:figI2}). But the same term produces 
also a virtual Dirac mass in a configuration $\tilde z_1(t) \neq z_1(t)$, according to 
\eqref{eq:figI3}. This Dirac mass is compensated by a different term of the expansion.
That is, the {\em negative} contribution coming from $\G(1,2)= (1,2)$ with $\bs_2=(-,+)$. To find this 
contribution, just let particle $1$ {\em ignore} the last collision with particle $2$ in 
Figure \eqref{eq:figI1}.

\smallskip
 
 We deduce now from Theorem \ref{thm:main} the validity of the BBGKY hierarchy for $L^1$
 measures:
\begin{cor} \label{cor:MS}
Given an initial measure on $\MM_N$ with density $W_0\in L^1(\MM_N)$ and
invariant for permutations of particles, let $\{f_j(t)\}_{j=1}^N$ be the family of time-evolved marginals as defined 
in \eqref{eq:Liou}-\eqref{eq:marg}. Then, the expansion \eqref{eq:BBGexp} holds for any $t>0$, 
almost everywhere in~$\MM_N$.
\end{cor}

To prove the corollary, let $\bz_N(t)$ be distributed according to $W(t)$ and consider
$\D_j(t)$ as a random probability measure over $\RRR^{6j}$. By Theorem \ref{thm:main},
the family of point processes $\{ \D_j(\cdot,\cdot) \}_{j=1}^N$ satisfies \eqref{eq:BBGexp}:
\bea
 \D_j (t) = && \sum_{n = 0}^{N-j} \alpha(N-j,n)  \int_0^t dt_1 \int_0^{t_1}dt_2\cdots\int_0^{t_{n-1}}dt_n \nn\\
&& {\cal S}^\e_j(t-t_1)\CC _{j+1}{\cal S}^\e_{j+1}(t_1-t_2)\cdots 
\CC _{j+n}{\cal S}^\e_{j+n}(t_n) \D_{j+n} \label{eq:Bdjt}
\eea
for any $\bz_N \in \MM_N^\star$. Thus, we just need to compute the expectation $\mathbb{E}$ of the 
above relation with respect to $W_0$. Observe that this amounts to apply $\int_{\MM_N}
d\bz_N W_0(\bz_N)$ in both sides.
Using \eqref{eq:Liou}-\eqref{eq:marg}, the symmetry in the exchange of particles and \eqref{micromart},
it easily follows that $\mathbb{E}\left[ \int_{\MM_j} \D_j (t) \,\phi\right] = \int_{\MM_j} f_j(t)\,\phi\;.$
On the other hand, in the assumptions of the corollary,
\bea
&&\int_{\MM_N\times A(t,\G,\bs_n)} d\bz_N d\bze_{j+n}\,\D_{j+n}(\bze_{j+n})\,\sum_{i=1}^{M} \,
\phi (\bze^{F,i}_j(t))W_0(\bz_N)\nn\\
&& = \sum_{\substack{i_1,\cdots,i_{j+n} \\ i_a \neq i_b}}\int_{\MM_N\times A(t,\G,\bs_n)} d\bz_N d\bze_{j+n}\,
\frac{ \prod_{s=1}^{j+n}   \d (\z_s- z_{i_s})}{N(N-1) \dots (N-j-n+1)} \,\sum_{i=1}^{M} \,
\phi (\bze^{F,i}_j(t))W_0(\bz_N)
\nn\\
&& = \int_{\MM_N\times A(t,\G,\bs_n)} d\bz_{N} d\bze_{j+n}\,
\prod_{s=1}^{j+n}   \d (\z_s- z_s) \,\sum_{i=1}^{M} \,
\phi (\bze^{F,i}_j(t))W_0(\bz_N) \nn\\
&& = \int_{A(t,\G,\bs_n)}  d\bze_{j+n}\,
\sum_{i=1}^{M} \,\phi (\bze^{F,i}_j(t))\,\int_{\MM_{j+n}} d\bz_{j+n}\prod_{s=1}^{j+n}\d (\z_s- z_s) \, f_{0,j+n}(\bz_{j+n}) \nn\\
&& = \int_{A(t,\G,\bs_n)}  d\bze_{j+n} \,f_{0,j+n}(\bze_{j+n})\,\sum_{i=1}^{M}\phi (\bze^{F,i}_j(t))\nn\\
&& =\e^{2n} \int d\bze_j\,d{\Lambda}\, \prod |B^\e_{\bs_n}|\,f_{0,j+n}(\bze_{j+n})\,\phi(\bze_j)\;,\nn
\eea
where in the last step we applied the change of variables \eqref{eq:eqmeas}.
Therefore, the expectation of the right hand side of \eqref{eq:Bdjt} represented in the weak
formulation \eqref{eq:Wfhe} is equal to the (integrated) expansion \eqref{eq:BBGexp},
i.e. \eqref{eq:WSfhe} holds for any bounded continuous $\phi$. This concludes the proof 
of the corollary.


\subsection {Microscopic series solution to the Enskog hierarchy} \label{sec:MSE}
\setcounter{equation}{0}    
\def\theequation{3.1.\arabic{equation}}

In this Section  we  consider, at a formal level, the Enskog hierarchy and its series solution, 
denoted by $\mu^E(t)$, for initial microscopic states $\mu_N(0) = \mu_N$ 
with $N\e^2 = \lambda^{-1}$ supported in $\bz_N \in \MM_N^*$, 
which reads (see \eqref {eq:fjexpE}, \eqref{eq:gjbzjt}, \eqref{eq:WfheE}):
\bea
&& \mu^E (t) = \sum_{n\geq 0}\lambda^{-n}\int_0^t dt_1 \int_0^{t_1} dt_2 \cdots \int_0^{t_{n-1}}dt_n \nn\\
&& \ \ \ \ \ \ \ \ \ \ \ \ \ \ \ \ \cdot{\cal S}_1(t-t_1)\CC_{2}{\cal S}_{2}(t_1-t_2)\cdots\CC_{1+n}{\cal S}_{1+n}(t_n) 
 \,  \mu_N ^{\otimes (1+n)} \nn\\
&& \ \ \ \ \ \ \ \ = \sum_{n=0}^\infty \sum _{\G (1,n)} \sum_{\bs_n}\lambda^{-n}
\int d {\Lambda}(\bt_n , \bo_n , \bv_{1,n}) \prod  B_{\bs_n} \,  \mu_N ^{\otimes (1+n)}(\bze^E(0)) \;,
\eea
or also:
\be
\int_{\RRR^6} \mu^E(t) \phi= 
\sum_{n = 0}^{\infty}  \sum_{\G(1,n)} \sum_{\bs_n} \prod_{r=1}^n \s_r \,  \lambda^{-n}\, \e ^{-2n}  
\int_{A^{E}(t,\G,\bs_n)}  d\bze^{E}_{1+n}\,\mu_N^{\otimes (1+n)}(\bze^{E}_{1+n})\,\phi (\bze^{E}_j(t))\;.
\label{eq:Enskwp}
\ee

The aim is to understand whether one may conclude that $\mu^E(t) = \mu_N(t)$. Since the expansion 
for $(\mu^E(t))^{\otimes j}$ satisfies the semigroup property (see \eqref{semigE}), we are allowed to 
proceed exactly as done in the previous section for the H-S hierarchy and focus on the single time 
interval in the partition \eqref{eq:PART}.

In doing so, we shall compare the result \eqref{eq:mainB}
with the right hand side of \eqref{eq:Enskwp}. Since we fix $N\e^2 = \lambda^{-1}$, the only difference
between the two is the use of the IFF instead of the EBF and, correspondingly, the different domains of integration
(remind also the discussion at the end of Section \ref{sec:hs2}). In particular, while the domains in 
\eqref{eq:Enskwp} cover all $\RRR^{6(1+n)}$, there holds in contrast $A(t,\G,\bs_n)\subset \MM_{j+n}$, which 
takes into account the non factorization of the initial state in \eqref{eq:mainB}. Of course,
outside $\MM_{j}$, the tensor product $\mu_N^{\otimes j} \neq 0$ because of the {\em contractions}
of type 
\be
\d(\z_1-z_i)\d(\z_2-z_i)\;.
\ee
In other words, \eqref{eq:ident} holds only over the phase space ``with holes'' $\MM_j$.

Observe now that the expansions in \eqref{eq:mainB} for $j=1$ and in \eqref{eq:Enskwp} 
appear to be identical when $t$ is so small that Properties 1 and 2 are satisfied in the single
time interval $(0,t)$\footnote{The main statement of \cite{Bog75} is in fact that the first equation of the
BBGKY hierarchy for empirical marginals reduces to the Boltzmann-Enskog equation. 
But this relies on the interpretation of the second marginal at the boundary via the product
formula \eqref{eq:ident}.}. Indeed in this case there are no recollisions, thus the interacting 
flow and the Enskog flow are the same. However, strictly speaking, the identity $\mu^E(t) = \mu_N(t)$
remains doubtful because the contractions in \eqref{eq:Enskwp} give contributions which are 
absent in \eqref{eq:mainB}. When considering such contributions, we face an ambiguity,
as explained by the example that follows.

\smallskip

Let us consider the simplest case $N=2$,  $N\e^2=\lambda^{-1}$ and assume that the two particles with initial configuration $\bz_2$  collide in the time interval $(0,t)$.

The term $n=0$ in \eqref{eq:Enskwp} is the free flow $\T^{E}_0 (\z_1,t) \mu_N$.
As before, this term cancels with the negative part of the term $n=1$, i.e. $\T^{E}_1 (\z_1,t) \mu_N$,
which is the contribution due to the trees
\be
\includegraphics[width=3in]{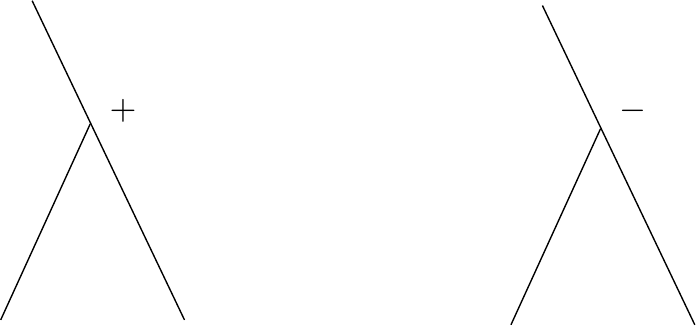}\nn
\ee
where the decorations $+$, $-$ correspond to $\s_1=+,-$ respectively.

Therefore we can write 
\be
\mu_N(t) = \T^{E}_0 (\z_1,t) \mu_N+\T^{E}_1 (\z_1,t) \mu_N\;. \label{eq:cn1n2E}
\ee
But there are other terms in \eqref{eq:Enskwp} which must be evaluated.

For instance, consider the tree
\be
\includegraphics[width=1.5in]{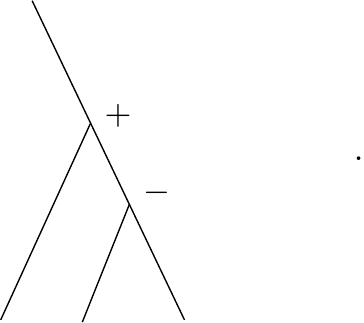}\nn
\ee
The associated contribution, integrated against $\phi$, gives
\bea
&& \lambda^{-2}\, \e ^{-4}  \int_{A^{E}}  d\z^{E}_{1}\,d\z^{E}_{2}\,d\z^{E}_{3}\,
\mu_N^{\otimes 3}(\z^{E}_{1},\z^{E}_{2},\z^{E}_{3})\,\phi (\z^{E}_1(t)) \label{eq:countn2}\\
&& = \frac{1}{N} \int_{A^{E}}  d\z^{E}_{1}\,d\z^{E}_{2}\,d\z^{E}_{3}\,\phi (\z^{E}_1(t))\nn\\
&&\,\,\,\,\,\, \cdot \Big[\d(\z^{E}_{1}-z_1)\,\d(\z^{E}_{2}-z_2)\,\d(\z^{E}_{3}-z_2)
+ \d(\z^{E}_{1}-z_2)\,\d(\z^{E}_{2}-z_1)\,\d(\z^{E}_{3}-z_1)\Big]\;,\nn
\eea
where $A^{E}$ is the image of the EBF-map $\z^{E}_1, \bt_2, \bo_2, \bv_{1,2} \rightarrow \bze^{E}_{3} (0)$,
namely the set defined by the condition: ``the free flow leads first particle $3$ and then particle $2$ at $\e-$distance 
from particle $1$''. Note that all the other contributions arising from the definition of $\mu_N^{\otimes 3}$
do vanish identically.

The dynamical content of \eqref{eq:countn2}, in terms of the EBF, is the following. 
Particle $1$ creates particles $2$ and particle $3$ {\em simultaneously}, at times 
$t_1=t_2$. Hence the deltas are evaluated on the border of $A^E$. One can be tempted to
interpret the numerical value of the above term as just $-\mu_N(t)$. This would 
cancel the contribution \eqref{eq:cn1n2E}. On the other hand, by a similar computation,
the value of $\mu_N(t)$ would be restored e.g. by 
\be
\includegraphics[width=1.5in]{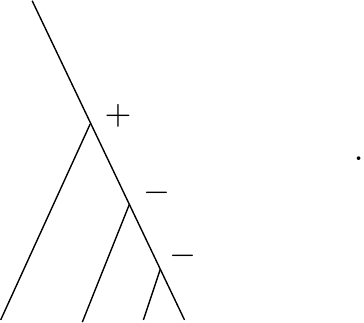}\nn
\ee

Furthermore, observe that the term 
\be
\includegraphics[width=0.8in]{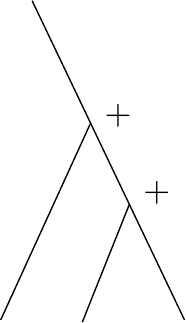}\nn
\ee
corresponds to a triple collision in the Enskog flow for which $\z^{E}_1(t)$ is not clearly defined.

\smallskip

The ambiguity we are dealing with cannot be easily solved by regularizing the initial $\d$-functions.
The obtained value of trees would depend on the regularization we choose. For instance, a 
symmetric regularization yields $-\frac 12 \mu_N(t)$ from \eqref{eq:countn2}, as a consequence
of the time ordering present in the definition of $A^{E}$. 
Moreover, families of trees yielding the same contribution with alternate
signs can be constructed, as in the example above, so that the resulting series is divergent 
(or at most converging in the Ces\`{a}ro sense). In conclusion, it is difficult to give a definite 
meaning to \eqref{eq:Enskwp} without further prescriptions.

It is not possible, however, to solve the problem by giving a prescription that simply eliminates
contractions\footnote{Or by considering the Enskog hierarchy \eqref{eq:WfheE} with correlated
initial data $g_{0,j} = \d_{j\leq N}\D_j$.}. 
These are indeed essential to reconstruct the H-S dynamics. In fact remind that, in the EBF, 
the only interactions are {\em creations} of particles (see Sec. \ref{sec:hs2}). 
Now consider for instance a case $N=3$ with three
or more interactions in the H-S dynamics, as in \eqref{eq:figI1}. In order to generate such a dynamics
in \eqref{eq:Enskwp}, we strictly need a term $\G(j,n)$ with $n>2$, which will have total number of particles
$j+n > 3$. Since $N=3$, this means that at least 2 particles occupy the same configuration. In other words,
recollisions in the H-S dynamics are replaced by creations and contractions in the EBF.

One way to eliminate the pathologies we have discussed for the simple example of a single collision,
is to consider a regularization based on the separation of times in the Duhamel series:
\bea
&& \mu^E (t) = \sum_{n\geq 0}\,\lambda^{-n}\,\lim_{\eta \to 0}\,
\int_0^t dt_1 \int_0^{t_1-\eta} dt_2 \cdots \int_0^{t_{n-1}-\eta}dt_n \nn\\
&& \ \ \ \ \ \ \ \ \ \ \ \ \ \ \ \ \cdot{\cal S}_1(t-t_1)\CC_{2}{\cal S}_{2}(t_1-t_2)\cdots\CC_{1+n}{\cal S}_{1+n}(t_n) 
 \,  \mu_N ^{\otimes (1+n)}\;. \label{renor}
\eea
With this definition, all the contraction terms discussed above are avoided.
In particular, the same computation of the proof of Theorem \ref{thm:main} for the case $j=1$ 
shows that, in the single time interval of the partition \eqref{eq:PART}, $\mu^{E}(t)= \mu_N(t)$ holds.
However, the regularized series does not converge to $\mu_N(t)$ for arbitrary times
(notice that the semigroup property fails).

\smallskip

We conclude with the remark that many singular solutions to the Boltzmann-Enskog equation which are not corresponding to the particle dynamics may be easily constructed. 
For instance, consider $N=2$ and the initial condition
\be
\frac 12 \Big[\d (\xi-x_1) \d(\eta -v_1) + \d (\xi-x_2) \d(\eta)\Big]
\ee
with support outside $\MM_2$, namely
$|x_1-x_2 | <\e\;.$ 
Then the unique solution to the equation is 
\be
 \frac 12 \Big[\d (\xi-x_1-v_1t ) \d(\eta -v_1) + \d (\xi-x_2) \d(\eta)\Big]\;,
\ee
because only the term $n=0$ is non-vanishing. Clearly, many other examples may be provided,
but it seems hard to construct a solution reproducing the physical dynamics, at least in the sense of 
\eqref{eq:WfheE}.

\bigskip
\bigskip
\bigskip

\ni {\bf Acknowledgments.} We are grateful to Maxime Hauray for interesting discussions on the subject.

\end{document}